\documentclass[12pt]{article}
\usepackage{epsfig}
\begin{document}

\begin {center}
{\Large An update on the Kappa}
\vskip 3mm
{D.V.~Bugg}
\vskip 2mm
{Queen Mary, University of London, London E1\,4NS, UK}
\end {center}

\begin{abstract}
New FOCUS data on $D^+ \to K^-\pi ^+\pi ^+$ are fitted to the
$\kappa$, together with earlier data from LASS, E791 and BES 2.
There is a clear low mass $K\pi$ peak due to the $\kappa$ pole.
Inclusion of the $I=3/2~K\pi$ amplitude gives only a marginal
improvement to the fit and almost no change to the $\kappa$ peak.
An improved formula for the $\kappa$ gives a better fit than that used
earlier.
The $\kappa$ pole moves to
$663 \pm 8(stat) \pm 34(syst) -i[329 \pm 5(stat) \pm 22(syst)]$ MeV.
The $K_0(1430)$ pole is at $1427 \pm 4(stat) \pm 13(stat)
-i[135 \pm 5(stat) \pm 20(syst)]$ MeV.
\end{abstract}

\vskip 2mm

\section {Introduction}
There are two objectives in this paper.
The first is to present a fit to new FOCUS data \cite {Focus}
and compare them with earlier E791 data \cite {E791}.
There are some systematic discrepancies between these two sets of
data, although their effects are minor.
Like E791, the FOCUS group has determined the $K^-\pi ^+$ S-wave
amplitude in magnitude and phase in 40 mass bins covering the whole
mass range available in $D^+ \to K^-\pi ^+\pi ^+$.
These data are fitted here.

The second objective is to introduce a necessary modification to the
formula used earlier to fit the $\kappa$.
Both $\sigma$ and $\kappa$ resonances have widths which are
strongly $s$-dependent, giving them unusual features.
This $s$-dependence originates directly from chiral symmetry breaking.
Hopefully, the formulae developed here will be useful to experimental
groups as direct replacements for the usual Breit-Wigner amplitudes
used to fit $J^P=0^+$ resonances such as $\kappa$ and $K_0(1430)$
in Dalitz plots.
The parametrisation for the $K\pi$ channel is highly convergent and
well determined in terms of just three parameters, one
of them related closely to $f_\pi ^2$.
Some allowance is also required for coupling to $K\eta$ and $K\eta '$,
but in the absence of direct experimental information on these
channels, this requires just one coupling constant for each channel.
If data for these channels become available, the present treatment
of $K\pi$ could be substituted also for $K\eta $ and $K\eta '$.

\section {Formulae}
The key point about both $\sigma$ and $\kappa$ is the Adler zero in the
elastic amplitude at $s \simeq m^2_K - m^2_\pi /2$ for the $\kappa$ and
at $s \simeq m^2_\pi /2$ for the $\sigma$.
These zeros are crucial features of chiral symmetry breaking.
The mechanism of this symmetry breaking is fully understood
\cite {Ribiero} \cite {Bicudo} \cite {Roberts} and is confirmed in
outline by Lattice QCD calculations \cite {Lattice}.
References [3-5] give illuminating discussions of the detailed
mechanism.
Pioneering work on the $\kappa$ and $\sigma$ was done by
Pelaez, Oset and Oller and collaborators \cite {Pelaez} \cite
{Oset} \cite {OO} \cite {Jamin}.

The consequence of the Adler zero is that the $K\pi$ amplitude rises
nearly linearly with $s$ near threshold.
In Ref. \cite {Comments}, the $\kappa$ amplitude was parametrised as
a Breit-Wigner amplitude with $s$-dependent width:
\begin {eqnarray} f_{el} &=& \frac {N(s)}{D(s)}
  = \frac {M\Gamma _{el}}{M^2 - s - iM\Gamma _{total}} \\
M\Gamma _{el} &=& \frac {s - s_A}{M^2 - s_A}
\exp (-\alpha k^2 ) g^2(K\pi ) \rho _{K\pi }(s);
\end {eqnarray}
here $s_A = 0.2367$ GeV$^2$ is the mean position of the Adler zero for
$K^+\pi ^-$ and $K^0\pi ^0$, and
$\rho (s)$ is Lorentz invariant phase space $2k/\sqrt {s}$, where
$k$ is centre of mass momentum.
Also $g$ are coupling constants and $\alpha$ is a fitted constant.
The exponential form factor in (2) accommodates the experimental fact
that $\Gamma _{el}$ gradually flattens off at large $s$.

There are two weaknesses in these formulae.
The important one is that they assume the $\kappa$ phase shift
eventually reaches $90^\circ$, although it was found in \cite
{Comments} that this was only at $\sim 3.3$ GeV, well above the mass
range of available data.
The FOCUS data, combined with earlier data, now reveal that the
$\kappa$ phase shift appears to reach only $\sim 55^\circ$ at 1.5
GeV and may never reach $90^\circ$.
It is straightforward to modify the formulae to cater for this
possibility.
The second weakness is that the mass and width of the pole itself are
far removed from $M$ and $\Gamma_{total}$ of Eqs. (1) and (2), for
reasons described in Section 4.
Formulae can be rewritten to be more closely related to the $\kappa$
pole itself.

A better form of the equations may be obtained by dividing both
numerator and denominator of eqn. (1) by $M^2$ and writing
\begin {eqnarray}
f_{el} = \frac {b_1(s - s_A) F_1(s) \rho _{K\pi}(s)}
{1 - s/M^2 - i \sum _j b_j(s - s_{A_j}) F_j(s) \rho _j(s)}.
\end {eqnarray}
For the $\kappa$, the term $s/M^2$ is small;
$b_j$ are constants and the summation in the denominator is over
$j=1-3$ for $K\pi$, $K\eta$ and $K\eta '$ channels in principle, though
the coupling to $K\eta $ turns out to be  insignificant.
Form factors $F_j$ are discussed in detail below.

Weinberg predicted the scattering lengths of pions from any
target \cite {Weinberg}; this sets the scale of chiral symmetry
breaking, hence $b_1$ in terms of $f_\pi ^2$.
Eq. (3) can be recast so as to expose the scattering length explicitly.
This requires a transformation in  $s$ of the denominator of (3).
The algebraic manipulations are to write
$1 - s/M^2 = 1 -As \simeq [1 - A(s - s_{thr})]/(1 + As_{thr})$
for small $A$, next
multiply top and bottom of (3) by $(1 + As_{thr})$ then replace all
$b_j$ by $B_j = b_j(1 + As_{thr})$; here $s_{thr}$ are evaluated at
the threshold of each channel $j$.
The result is
\begin {equation}
f_{el} = \frac {N(s)}{D(s)} = \frac {B_1(s - s_A)F_{el} \rho _{K\pi}(s)}
{1 - A(s - s_{thr}) - i \sum _j B_j(s - s_{A_j}) F_j(s) \rho _j(s)}.
\end {equation}
If $A$ is positive, $Re ~D$ eventually goes to zero and the elastic
phase shift reaches $90^\circ$.
The case where $A$ is negative accomodates the possibility that the
phase shift does not reach $90^\circ$.
The denominator however still allows a $\kappa $ pole.
The scattering length $a$ is
\begin {equation}
a = 2B_1(s_{thr}-s_A)/\sqrt{s_{thr}}.
\end {equation}

There is an alternative way of viewing this formula.
The factor $1 -A(s - s_{thr})$ in the denominator of Eq. (4)
may be regarded as an empirical form which exhibits the scattering
length explicitly and parametrises successfully the real part
of the amplitude in a region close to threshold and the Adler zero.
For elastic scattering, a form factor allowing a controlled departure
of the numerator $N(s)$ from linearity is
$F_1 = \exp (-\alpha _1k_1^2)$, where $\alpha_1$ is a fitted parameter.
Note that the exponential dependence of the form factor combines
both a conventional form factor for the radius of interaction
and an empirical departure of the numerator from linearity;
values of $\alpha_1$ and $A$ are accurately determined by FOCUS and
E791 data.

The $K\eta$ channel turns out to be negligible; for $K\eta '$, the
Adler zero is at $s = m^2_{\eta '} - m^2_K/2$.
The form factor for $K\eta '$ is not known, because of lack of data
for this channel.
Above its threshold, the value $\alpha_j = 4.5$ (GeV/c)$^{-2}$ is
adopted from a wide range of Crystal Barrel and BES II data; this value
corresponds to a reasonable radius of interaction 0.72 fm.
Above the $K\eta '$ threshold, small changes in $\alpha_j$ may be taken
up by small alterations to parameters fitted to $K_0(1430)$ and
$K_0(1950)$ and the fit to data changes by less than the errors.

Below the inelastic threshold, the Flatt\' e prescription is adopted,
continuing $\rho$ analytically: $\rho \to i|\rho |$ \cite {Flatte}.
However, $|\rho|$ increases below threshold and requires a cut-off;
otherwise, a myriad of open channels at high mass dominate $K\pi$
elastic scattering at low mass.
Such a prescription would obviously be unphysical.
With the accuracy of present data, any reasonable cut-off will do, and
the simple one
\begin {equation} F_j = \exp [-\alpha_j|k^2_j|]
\end {equation}
fits data adequately.
The best source of information on sub-threshold form factors comes from
Kloe data on $\phi \to \gamma \pi ^0 \pi ^0$, where a similar cut-off
is required for the sub-threshold $\sigma \to KK$ amplitude \cite
{Recon1}.
Rather a sharp cut-off is required for those data and optimises with
$\alpha _j = 8.4$ (GeV/c)$^{-2}$, though a value as low as half this
is acceptable.
For present data, a similar form factor is definitely required.
If a value $\alpha _j < 2$ (GeV/c)$^{-2}$ is used, the
$K_0(1430) \to K\eta '$ amplitude near the $K\pi$ elastic threshold
becomes unreasonably large.
There is a weak optimum at $\alpha _j = 4.5$ (GeV/c)$^{-2}$.
With this value, the sub-threshold contribution rises rapidly from
the $K\eta '$ threshold to a $K\pi$ mass of 1400 MeV and thereafter
varies slowly and smoothly down to the $K\pi$ threshold.
Any alteration to this value of $\alpha _j$ is absorbed by small
changes to the fitted parameters of $K_0(1430)$ and small changes
to fitted values of $\alpha _1$ and $B_j$.
The extrapolation of the amplitude to the pole is stable: the real
and imaginary parts of the pole position change
by 20--30 MeV for variations of $\alpha _j$ in the range
3.5--6.5 (GeV/c)$^{-2}$.
This will be taken as a systematic error on the pole position.
[If data one day become available on the form factor for $K\eta '$
above threshold, it will be possible to calculate the sub-threshold
continuation from a dispersion relation like Eq. (8) discussed below,
but presently this is academic.]

For production processes such as $D \to (K\pi )\pi$, the denominator
$D(s)$ of the amplitude must be identical to that of elastic scattering
by Watson's theorem \cite {Watson}.
However, the numerator can be (and is) very different.
Data on $J/\Psi \to \omega \pi ^+\pi ^-$ \cite {wpp} and
$K^+\pi ^- K^-\pi ^+$ \cite {KKpp} \cite {DVB}
and present data on $D \to (K\pi )\pi$ may be fitted accurately
taking $N(s)$ as a constant.
The result is that $\sigma$ and $\kappa $ poles appear clearly in
production data, but are obscured in elastic scattering by the nearby
Adler zero in $N_{el}(s)$.
Precise theoretical work on elastic scattering using the Roy equations
does however reveal $\sigma$ and $\kappa$ poles clearly \cite {Caprini}
\cite {Descotes}.

The Adler zero is a property of the full $K\pi$ elastic amplitude
and therefore needs to be included into the widths of $K_0(1430)$
and $K_0(1950)$.
This is done using Eqs. (1) and (2) for each of $K_0(1430)$ and
$K_0(1950)$; the exponential form factor is taken to be the same as for
the $\kappa$.
Although $K_0(1430)$ may be fitted with a Breit-Wigner
amplitude of constant width, the resulting contribution to the
scattering length is far too large to agree with Weinberg's prediction.

In order to satisfy unitarity for LASS data, the full S-matrices of
$\kappa$, $K_0(1430)$, $K_0(1950)$ and the $K\pi ~I=3/2$
amplitude are multiplied.
This prescription is not unique, but gives a good fit to the
interference between $\kappa$ and $K_0(1430)$.

The FOCUS, E791 and BES 2 groups have all determined the magnitude
and phase of the $K\pi$ S-wave with respect to other strong components
in the data.
To fit production data, the isobar model is used, with a complex
coupling constant in the numerator of each amplitude instead of
$M\Gamma _{el}$, specifically
$\Lambda \exp(i\phi)/(M^2 - s - iM \Gamma _{tot})$.
The first two sets of data require a phase difference between
$\kappa$ and $K_0(1430)$ $\sim 75^\circ$ different to elastic
scattering.

One alternative procedure for fitting the data has been explored,
but in practice turns out to give no better result than the Flatt\' e
prescription, though it does provide a cross-check.
In principle, more complete formulae for a Breit-Wigner resonance
are \cite {Sync}
\begin {eqnarray}
f_{el}  &= &\frac {M\Gamma _{el}}{M^2 - s - m(s) - iM\Gamma _{total}
(s)}, \\
m(s) &=& \frac {M^2 - s}{\pi } \int \frac {ds'~\Gamma _{total} (s')}
{(s' - s)(M^2 - s')}.
\end {eqnarray}
The scattering amplitude is an analytic function of $s$.
Any $s$-dependence in $M\Gamma_{total}$ gives rise to a corresponding
contribution to the real part of the denominator via $m(s)$.
The rapid opening of the $K\eta '$ threshold generates
a sharp spike in $m(s)$ precisely at the $K\eta '$ threshold.
A subtraction in (8) at mass $M$ improves the convergence of the
integral.

The problem which arises is that this spike depends on the mass
resolution of the experiments.
Without an accurate knowledge of mass resolution and possible
variations over the Dalitz plot, inclusion of $m(s)$ is impractical
for $K\eta$ and $K\eta '$ channels, though satisfactory for the
$K\pi$ channel.
The Flatt\' e formula only partially reproduces the effect of $m(s)$.
However, in practice, it allows small adjustments of $M$ and
$g^2_{K\eta '}/g^2_{K\pi }$ which fudge the effect of $m(s)$ by
optimising the fit to data.
Readers interested in the effect of $m(s)$ are referred to
Ref. \cite {a0980}, where it has been included into fits to
$a_0(980)$ in Crystal Barrel data.
In that case, the mass resolution is known accurately and included.
It smears out the spike at the $KK$ threshold seriously, as shown in
Fig. 6(b) of that paper.
Even then, the dispersion integral for $m(s)$ is sensitive to the form
factor, and the height of the spike needs to be fitted empirically.
Here, the dispersive correction will be used for the
$K\pi$ channel so as to assess possible systematic errors in
parametrising that channel.

\section {The fit to data}
\begin{figure} [htb]
\begin{center}
\epsfig{file=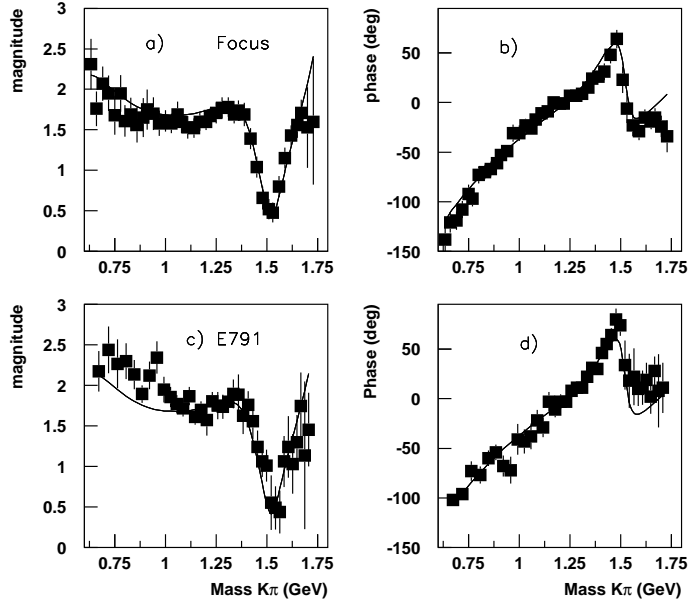,width=10cm} \
\vskip -6mm
\caption{The result of the combined fit for (a) magnitudes and (b)
phases  of FOCUS amplitudes; (c) and (d) the corresponding fit to E791
data}
\end{center}
\end{figure}
\begin{figure} [htb]
\begin{center}
\epsfig{file=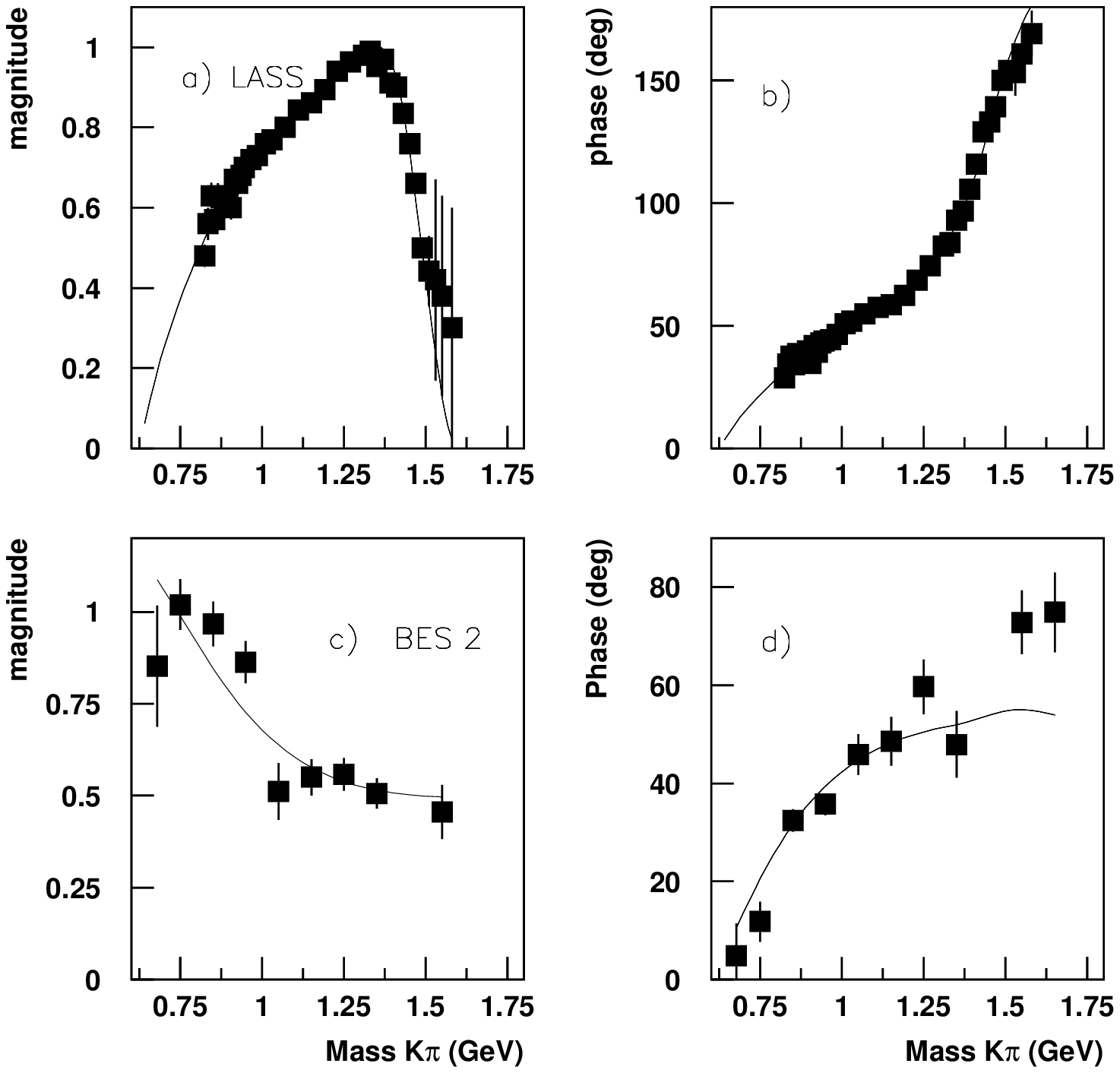,width=10cm} \
\vskip -6mm
\caption{The result of the combined fit for (a) magnitudes and (b)
phases of LASS amplitudes; (c) and (d) fits to BES 2 data.}
\end{center}
\end{figure}
An important feature of present work is that data from LASS for
elastic scattering \cite {Aston}, FOCUS \cite {Focus} and
E791 \cite {E791} and BES 2 \cite {DVB} are fitted simultaneously
with consistent parameters.
This reveals the strengths and weaknesses of each individual set of
data.
In BES 2 data, there is a strong $K_0(1430)$ peak, together with a
known smaller contribution from the narrower $K_2(1430)$.
This peak is included in the present fit since it gives the best
determination of parameters for $K_0(1430)$.

The E791 data were published including a form factor in $N(s)$ for
production, but it was subsequently shown \cite {DVBE791} that the fit
optimises with this form factor set to 1.
That correction has been applied to data fitted here.
FOCUS also take the form factor for production to be 1.
Figs. 1(a)-(d) show the fit to FOCUS and E791
data.
One sees some modest systematic discrepancies between them.
The fit to either may be improved by omitting the other, but
both are included in the final fit.
Values of $\chi ^2$ will be given for a variety of combinations
in Table 2;
$\chi ^2$ is calculated directly from FOCUS and E791 tabulations,
combining their statistical and systematic errors in quadrature.

Figs. 2(a) and (b) show the fit to LASS magnitudes and phases.
It is well known that these data stray slightly outside the unitarity
circle above 1.5 GeV, so errors in this region have been increased
until an average $\chi ^2$ of 1 is achieved; this refinement has no
significant effect on the fit.
Figs. 2(c) and (d) show the fit to BES 2 data.
The phases of two high mass points now look high; however, since they
are direct measurements of phase in the data, they are retained in the
fit.
The fit to the 1430 MeV peak in BES 2 data is shown in Fig. 3(a).
\begin{figure} [htb]
\begin{center}
\epsfig{file=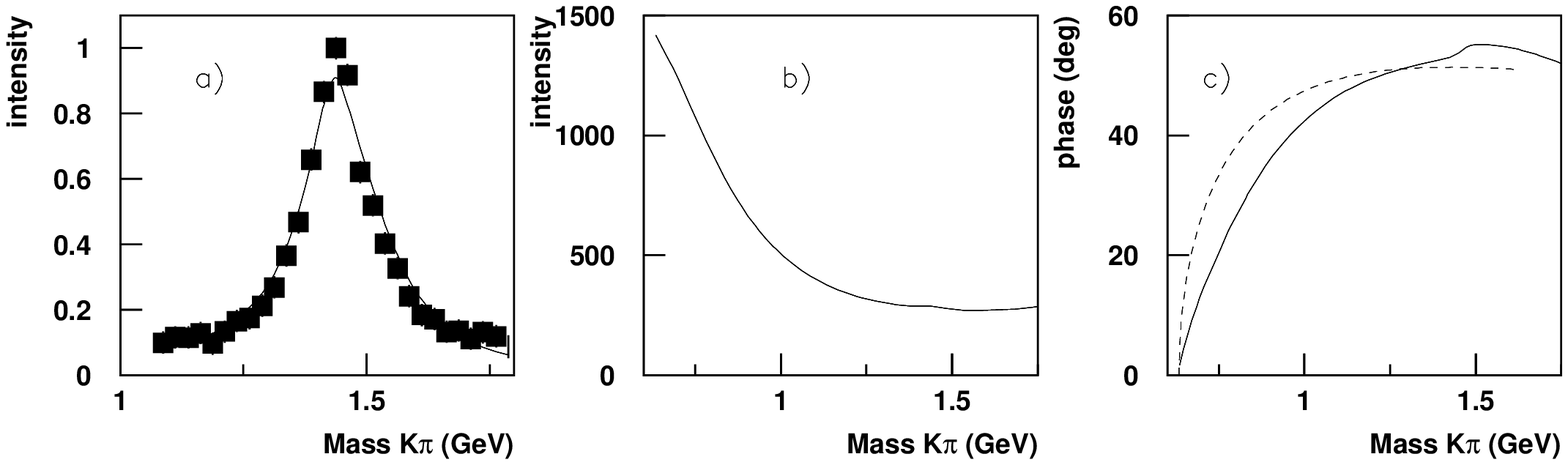,width=15cm} \
\vskip -6mm
\caption {The fit to (a) the intensity of the 1430 MeV peak
in BES 2 data; (b) intensity and (c) phase of the fitted $\kappa$
amplitude. In (c), the dashed curve shows the shape of the LASS
effective range formula.}
\end{center}
\end{figure}

The fit to all data requires small contributions from $K_0(1950)$.
Its parameters have been allowed to vary by $1\sigma$ from averages
quoted by the Particle Data Group \cite {PDG}.
They finish at $M = 1967.4$ MeV, $\Gamma _{EL} = 115$ MeV, $\Gamma
_{total} = 376$ MeV, but these changes have insignificant effect on
parameters fitted to $\kappa$ and $K_0(1450)$.

There is one quite large change in the fit to $K_0(1430)$ compared to
that reported in Ref. \cite {DVBE791}.
The ratio $g^2_{K\eta '}/g^2_{K\pi}$ reported there was 1.15.
This now falls to $0.62 \pm 0.06$.
This change may be traced directly to the improvement in Eqs. (3-5)
over Eqs. (1-2), used earlier.
The latter accommodated a $K\pi$ phase of $\sim 72^\circ$ in BES 2 data,
but forced a large value of $g^2_{K\eta '}$ for $K_0(1430)$ via  the
sub-threshold contribution of the $K\eta'$ channel. However, the
earlier fit was visibly not perfect from 1.15 to 1.35 GeV because this
contribution was too large. That problem now disappears. With the new
formulae, the phase shift of the $\kappa$ peaks at $55^\circ$ at 1530
MeV, though any fall between there and the end of the mass range is
within statistics. The magnitude and phase of the $\kappa$ contribution
are displayed in Figs. 3(b) and (c). The vertical scale for Fig. 3(b)
is arbitrary.

Inclusion of the coupling of $\kappa$ to $K\eta$ improves $\chi^2$ by
only 4.6, i.e. $2\sigma$.
It also tends to destabilise the fit because of interferences between
the sub-threshold $K\eta$ component and $K\pi$.
It is therefore omitted from the final fit.
There is now evidence for a component due to $\kappa \to K\eta '$.
It optimises at $g^2(\eta'K)/g^2(K\pi) = 0.085 \pm 0.020$ and improves
the fit by 14.6, i.e $3.8\sigma$.
It is not strongly correlated with the contribution from
$\kappa \to K\eta '$.
The small structure in the $\kappa$ phase near 1.5 GeV is due to
coupling to $K\eta '$.
Direct data on $K\eta'$ would be valuable to improve the determination
of the coupling of $\kappa$ and $K_0(1430)$  to this channel.

\subsection {Possible effects of the $K\pi~I = 3/2$ amplitude}
This amplitude has been parametrised by Pennington.
His formula for the K-matrix, quoted in Ref. \cite {FOCUS2}, will be
repeated here for completeness:
\begin {equation}
K_{3/2} = \frac {(s - 0.27)}{s_{norm}}
(-0.22147 + 0.026637\bar s -0.00092057 \bar s^2),
\end {equation}
where $\bar s = s/s_{norm} -1$,
$s_{norm} = (m_K^2 + m^2_\pi)$ and $s$ is in GeV$^2$.
Inclusion of this amplitude improves $\chi^2$ only by 7.7.
This is an indication that the data agree well with both the
magnitude and the phase of the broad $\kappa$ amplitude.
The improvement is not concentrated in any particular
mass region or data set, but is fragmented into small
improvements distributed almost randomly.
What happens in the fit is that destructive interference
develops between $\kappa$ and the $I=3/2$ amplitude.
These are familar symptoms that two broad amplitudes are
combining to  fit minor defects in the data.

There is a physics reason why the contribution from repulsive
amplitudes should be small.
For attractive interactions (particularly resonances), the
wave function is sucked into small radii, $r$; for repulsive
interactions, it is repelled to large $r$ by the potential
barrier and therefore reduced in magnitude.
There is evidence that $J/\Psi$  and $D$ decays involve
short-range interactions.
This comes from the absence of form factors in their decay
processes.
From the earlier analysis of E791 data, the interaction was
found to have an RMS radius $<0.38$ fm with $95\%$ confidence.

In an earlier FOCUS publication on $D^+ \to K^-\pi ^+\pi ^+$,
the repulsive $I = 3/2~K \pi$ amplitude was included, but
led to massive destructive interference with the $\kappa$.
In the earlier FOCUS analysis, the Adler zero was not
included into the $\kappa $ amplitude, which was parametrised by a
broad Breit-Wigner resonance and a constant interfering background.
With three broad amplitudes, destructive interferences are
able to patch up minor defects all over the Dalitz plot.
Data on other charge combinations could identify production
of $I=3/2$ $K\pi$ and $I=2~\pi \pi$ amplitudes.

The Cleo C collaboration has also presented an analysis of
$D^+ \to K^-\pi^+\pi ^+$ where they include the $I=2$ $\pi \pi$
amplitude \cite {extra}.
They present three solutions, all of which contain huge $I=2$
amplitudes with an intensity a factor $\sim 30$ larger than
$K^*(980)\pi$.
This is clearly symptomatic of large destructive interferences with
the $\kappa$ amplitude, since the $K^*(890)$ is clearly visible by
eye in the Dalitz plot and $K\pi$ mass projection.
Their second solution is made with an amplitude derived directly from
a complex pole at $M=706.0 \pm 1.8(stat) \pm 22.8(syst)
-i(319.4 \pm 2.2 (stat) \pm 20.2(syst)$ MeV.
In view of the huge interference with the $I=2$ amplitude, it is
difficult to know how reliable this result is.

\begin{table}[htb]
\begin {center}
\begin{tabular}{ccc}
\hline
   & Parameter & Value  \\\hline
$\kappa$  & $A$        & $-0.080 \pm 0.062$ \\
          & $B_1$      & $2.528 \pm 0.089$ \\
          & $\alpha_1$ & $0.566 \pm 0.056$ \\
          & $g^2_{K\eta '}/g^2_{K\pi }$ & $0.085 \pm 0.020$ \\\hline
$K_0(1430)$ & M        & $1.479 \pm 0.004$ \\
            & $g^2_{K\pi}$     &  $0.284 \pm 0.012$ \\
            & $g^2_{K\eta '}/g^2_{K\pi}$ & $0.62 \pm 0.06$ \\\hline
\end{tabular}
\caption{Parameters fitted to $\kappa$ and $K_0(1430)$ in units of GeV}
\end{center}
\end{table}

\subsection {Pole positions}
Table 1 gives fitted parameters for $\kappa$ and $K_0(1430)$
in units of GeV.
Pole positions are remarkably stable if individual sets of
data are dropped from the fit, or even pairs of sets.
As an illustration, Table 2 shows values when each of the sets of
data listed in column 1 are dropped.
There are 14 fitted parameters: the 7 listed in Table 1, 6 for three
complex coupling constants fitted to $\kappa$, $K_0(1430)$ and
$K_0(1950)$ in $D \to K\pi \pi$, and one scale factor for
the magnitude fitted to $K_0(1430)$ in BES 2 data.
The total $\chi ^2$ is 349.4 for 272 degrees of freedom.
Errors quoted in Table 1 for each parameter correspond to changes
in $\chi ^2 $ of 1.28, the mean value per degree of freedom.
However, there are some correlations between parameters, with the
result that the quoted errors overestimate somewhat the changes
expected  in Table 2.
The final two columns in Table 2 show changes in $\chi^2 $ when each
set of data is dropped and also the number of points dropped.
\begin{table}[htb]
\begin {center}
\begin{tabular}{ccccc}
\hline
Dropped          & $\kappa$  &  $K_0(1430)$ & $\Delta \chi^2$ & points
\\\hline
LASS phases      & 661 -i325 & 1432-i127 & 34.9 & 41 \\
LASS magnitudes  & 663 -i334 & 1435-i140 & 44.9 & 41 \\
FOCUS magnitudes & 663 -i329 & 1427-i137 & 41.7 & 40 \\
FOCUS phases     & 660 -i327 & 1427-i129 & 98.2 & 40 \\
E791 magnitudes  & 661 -i328 & 1426-i135 & 73.2 & 38 \\
E791 phases      & 663 -i329 & 1424-i139 & 48.2 & 38 \\
BES $K_0(1430)$  & 675 -i341 & 1431-i144 & 54.0 & 29 \\
BES magnitudes   & 640 -i321 & 1425-i132 & 21.3 &  9 \\
BES phases       & 657 -i323 & 1426-i133 & 27.4 & 10 \\
FOCUS all        & 660 -i328 & 1425-i132 & 158.4& 80 \\
E791 all         & 662 -i327 & 1424-i140 & 126.4& 76 \\\hline
\end{tabular}
\caption{Pole positions in MeV when individual sets of
data are dropped; the last two columns show the change in $\chi ^2$ of
the fit and the number of points dropped.}
\end{center}
\end{table}

Using all sets of data, the $\ K_0(1430)$ pole position is
\begin {equation}
M = 1427 \pm 4(stat) \pm 13 (syst) -
i[135 \pm 5(stat) \pm 20 (syst)]~{\rm MeV}.
\end {equation}
The systematic error is compounded from the worst  case
and from uncertainties in relative contributions of
$K_2(1430)$ and $K_0(1430)$ (and their relative phase) to BES 2
data on the 1430 MeV peak.
For the $\kappa$, the pole position is
\begin {equation}
M = 663 \pm 8(stat) \pm 34 (syst) -
i[329 \pm 5(stat) \pm 22 (syst)]~{\rm MeV}.
\end {equation}
Values for both $\kappa$ and $K_0(1430)$  supersede earlier
determinations in Ref. \cite {DVBE791} because of the improvement in
the formulae for the $\kappa$.
For the latter, further systematic errors are included to cover
uncertainties in the extrapolation to the pole,
estimated by changing form factors and a possible small contribution
from the $K\eta$ channel.
The $\kappa$ pole has moved from $750^{+30}_{-55}-i{342\pm 60}$
MeV \cite {DVBE791} by more than the quoted systematic error on mass
because of the improvement in the fitting formulae.
A systematic discrepancy with LASS data has been cured.
There is no apparent need to increase the flexibility in the fitting
formulae, but it is difficult to estimate systematic effects this might
have on the $\kappa$ pole.

The scattering length $a$ for the $\kappa$ component is given by Eq.
(5) except for a small correction from  the sub-threshold contribution
of $K\eta '$ via the denominator of  Eq. (4).
For the optimum fit, the $\kappa$ contributes $0.1860/m_\pi$ to the
scattering length, $K_0(1430)$ contributes $0.0086/m_\pi$ and
$K_0(1950)$ contributes $0.0002/m_\pi$, i.e. a total of
$(0.1950 \pm 0.0060)/m_\pi$.
Weinberg \cite {Weinberg} predicts $0.172/m_\pi$ and
corrections from Chiral Perturbation Theory to order $p^4$ increase
this to $(0.19 \pm 0.02)/m_\pi$ \cite {Chpt}.
Experiment and prediction are now in good agreement and experiment
is now better than theory.

The $\kappa$ pole position also now agrees well with the prediction of
Descotes-Genon and Moussallam \cite {Descotes} from the Roy equations,
$658 \pm 13 - i(278 \pm 12)$ MeV.
They considered only the mass range up to 1 GeV and omitted the
$K\eta '$ amplitude and its sub-threshold contribution.

For completeness, Table 3 shows additional parameters used in fitting
the production data of FOCUS and E791.
The experimental groups use $K^*(980)$ as a reference amplitude.
Table 3 therefore refers to parameters fitting their tabulated
amplitudes for the $K\pi$ S-wave.
Table 3 also  includes the normalisation factor $\Lambda _{norm}$
which scales the BES 2 data for $K_0(1430)$.
\begin{table}[htb]
\begin {center}
\begin{tabular}{ccccc}
\hline
Parameter          & Value
\\\hline
$\Lambda _{\kappa}$ & $2.98 \pm 0.09$ \\
$\Lambda _{1430}$   & $-0.72 \pm 0.04$ \\
$\Lambda _{1950}$   & $-3.16 \pm 0.07$ \\
$\phi _{\kappa}$    & $130.9 \pm 2.3$ \\
$\phi _{1430}$      & $53.9  \pm 2.6$ \\
$\phi _{1950}$      & $26.6 \pm 4.1$  \\
$\Lambda _{norm}$   & $0.925 \pm 0.028$ \\\hline
\end{tabular}
\caption{Additional parameters fitted to production data.}
\end{center}
\end{table}

\subsection {A check using the dispersive correction $m(s)$}
The $s$-dependence of $m(s)$ derived purely from the $K\pi$
component of $\Gamma_{total}$ is shown in Fig. 4, after a subtraction
at the $K\pi$ threshold.
The fit to data including $m(s)$ is almost indistinguishable from that
of Figs 1 and 2.
To check the scattering length, it is necessary to fit $m(s)$ to terms
in $k$ and $k^3$.
The derived scattering length is $(0.201 \pm 0.007)/m_\pi$, agreeing
with the result quoted above within the error.
One should note that the dispersion integral for $m(s)$ involves an
integral from threshold to infinity and is therefore sensitive to
assumptions about the behaviour of the $\kappa$ amplitude above the
available mass range.
There could be contributions from further inelastic channels such as
$\kappa \sigma$.
The $K\pi$ amplitude itself `knows' about such contributions, so the
determination of the scattering length direct from fitted amplitudes
at low mass is likely to be the more reliable.

\begin{figure} [t]
\begin{center}
\epsfig{file=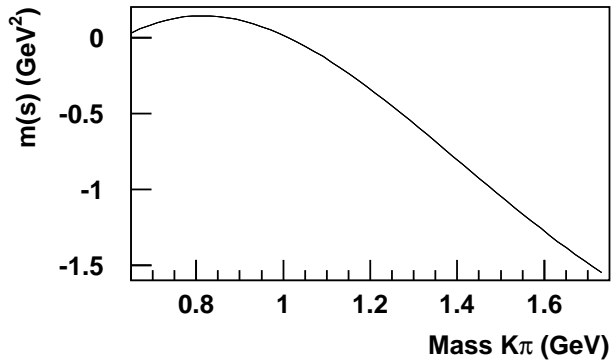,width=10cm}\
\vskip -6mm
\caption{m(s) evaluated from Eq. (8) with a subtraction at
the $K\pi$ threshold }
\end{center}
\end{figure}

\subsection {How the $\kappa$ amplitude varies off the real $s$-axis}
The $K\pi$ and $\pi \pi$ amplitudes are analytic functions of $s$.
From the Cauchy-Riemann relations,
\begin {eqnarray}
\partial (Re \,f)/\partial (Re \,s) &=&
\partial (Im \,f)/\partial (Im \,s) \\
\partial (Im \,f)/\partial (Re \,s)  &=& -\partial (Re \,f)/d(Im \,s),
\end {eqnarray}
there is a rapid variation of both real and
imaginary parts of $f$ with $Im \,s$.
This leads to a rotation of the phase of the amplitude with $Im \,s$.
On the real $s$-axis, unitarity requires that the phase is zero at
threshold, but it moves negative with increasing negative values of
$Im \,s$.

This is illustrated in Fig. 5 for four values of $Im \,s$.
Note the the phase below the pole moves from a small value
for the full curve at $Im~m = -0.2$ GeV to a large negative
value for $Im~ m = -0.32$ GeV, very close to the pole.
This explains the curious result that there can be a pole with a real
part close to zero at threshold.
Oller \cite {Oller} has drawn attention to this point in a somewhat
different way.
On the real $s$-axis, what one sees is similar to the upper part of
the pole, but rotated in phase.
This explains why the phase for real $s$ does not reach
$90^\circ$.
It reinforces the fact that $M$ and $\Gamma$ of Eqs. (1) and (2) are
remote from those of the pole.
\begin{figure} [t]
\begin{center}
\epsfig{file=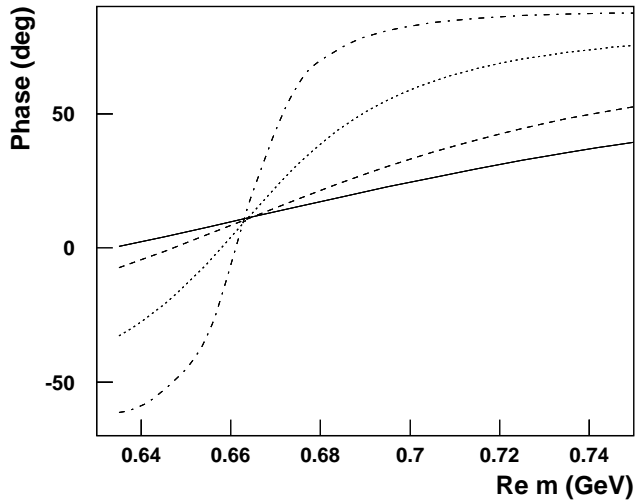,width=10cm}\
\vskip -6mm
\caption{The phase variation near the pole v. mass,
for Im m = -0.2 GeV (full curve), -0.25 GeV (dashed), -0.30 GeV
(dotted) and -0.32 (chain curve).}
\end{center}
\end{figure}

\section {Discussion of results}
The key features emerging from current data are (i) the $\kappa$ peak
near threshold, Fig. 3(b), and (ii) the precise form of the phase shift
near threshold, Fig. 3(c).
The FOCUS collaboration \cite {Focus} illustrate in their Fig. 8 the
fact that their measured phases near threshold lie distinctly below
those obtained from a conventional effective range form.
This is of course due to the nearby Adler zero, which is absent from
the usual effective range form.
The dashed curve in Fig. 3(c) illustrates the LASS effective range
formula, which is more curved near threshold.

Insight into the nature of $\sigma$ and $\kappa$ is provided by the
model of confinement constructed by van Beveren and Rupp \cite
{Lisbon}.
In this simple model, confinement is approximated by a harmonic
oscillator potential (which can be solved algebraically), matched to
plane wave states at a boundary.
The way the model is constructed, it approximately reproduces the
effect of the Adler zero.
With the boundary at 0.7 fm, the model reproduces quite well the
parameters of all of $\sigma$, $\kappa$, $a_0(980)$ and $f_0(980)$
with a single universal coupling constant.
A reasonable phase angle is needed between $u\bar u$ and $s\bar s$ to
reproduce $f_0(980)$ and $\sigma$ amplitudes near 1 GeV.
The $f_0(980)$  and $a_0(980)$ are locked to the $KK$ threshold by the
sharp cusp in  $m(s)$ due to the opening of this threshold.
The $a_0(980)$ does not appear at the $\eta \pi$ threshold because of
the Adler zero close to this threshold.
It is noteworthy that this model was the first (in 1986) to reproduce
the lowest scalar nonet \cite {First}, with the title: `A low-lying
scalar meson nonet in a unitarised meson model'.

In this model, $\sigma$, $\kappa$, $a_0(980)$ and $f_0(980)$ are a
nonet of continuum states coupled to the confining potential at its
boundary.
They are meson-meson states at large $r$, coupled to $q\bar q$
components within the confining potential.
Doubtless the model could be improved by adding meson exchanges at
large $r$.
In more detail, it is also possible that diquark interactions play a
role via coloured configurations, as proposed by Jaffe \cite {Jaffe}.

There is a clear analogy between $\sigma$ and $\kappa$ and the
weak interaction because the amplitude rises linearly with
$s$ near threshold.
The scale of the electroweak interaction is set by the masses of
$W$ and $Z$.
If the Higgs boson appears as a broad pole like $\sigma$ and $\kappa$,
dispersive effects due to opening of $WW$, $WZ$, $ZZ$ and $ t\bar b$
thresholds will play an important role \cite {Alboteanu}.

I am grateful for discussions with Profs. G. Rupp, E. van Beveren,
P. Bicudo and J. Ribiero over a period of many years.

\end {document}